\documentclass[unsortedaddress,preprint,aps,showpacs]{revtex4}
\usepackage{graphicx}
\usepackage{dcolumn}
\usepackage{amsmath}
\usepackage{bm}
\usepackage{color}
\usepackage{hyperref}
\newcommand{\be}{\begin{equation}}
\newcommand{\ee}{\end{equation}}
\newcommand{\ba}{\begin{eqnarray}}
\newcommand{\ea}{\end{eqnarray}}
\begin{document}
\title{Hexatic phase and water-like anomalies in a two-dimensional fluid of particles with a weakly softened core\footnote{This article appeared in \emph{The Journal of Chemical Physics} {\bf 137}, 104503 (2012) and may be found at \url{http://link.aip.org/link/?jcp/137/104503}. Copyright (2012) American Institute of Physics. This article may be downloaded for personal use only. Any other use requires prior permission of the authors and the American Institute of Physics.}}
\author{Santi Prestipino\footnote{Email: santi.prestipino@unime.it}}
\affiliation{Universit\`a degli Studi di Messina, Dipartimento di Fisica,
Contrada Papardo, 98166 Messina, Italy}
\author{Franz Saija\footnote{Corresponding author; email: saija@me.cnr.it}}
\affiliation{CNR-IPCF, Viale Ferdinando Stagno d'Alcontres 37, 98158 Messina, Italy}
\author{Paolo V. Giaquinta\footnote{Email: paolo.giaquinta@unime.it}}
\affiliation{Universit\`a degli Studi di Messina, Dipartimento di Fisica,
Contrada Papardo, 98166 Messina, Italy}
\begin{abstract}
We study a two-dimensional fluid of particles interacting through a
spherically-symmetric and marginally soft two-body repulsion. This model
can exist in three different crystal phases, one of them with square
symmetry and the other two triangular.
We show that, while the triangular solids first melt into
a hexatic fluid, the square solid is directly transformed on heating into
an isotropic fluid through a first-order transition, with no intermediate
tetratic phase. In the low-pressure triangular and square
crystals melting is reentrant provided the temperature is not too low,
but without the necessity of two competing nearest-neighbor
distances over a range of pressures. A whole spectrum of water-like fluid
anomalies completes the picture for this model potential.
\end{abstract}
\pacs{61.20.Ja, 64.70.D-, 65.20.-w, 68.60.-p}
\maketitle

\section{Introduction}

When confined to two-dimensional (2D) space, condensed matter behaves
differently than in three dimensions (3D). A striking
example is the physics of the Kosterlitz-Thouless transition in superfluid
films, not to mention the phenomenon of high-temperature superconductivity
in the cuprates, which is intimately
related to the motion of electrons within weakly-coupled copper-oxide layers.
Another example is provided by 2D crystals, where thermal fluctuations are
so strong as to rule out long-range translational (but not orientational)
order for non-zero temperatures, leaving it open for the possibility of
unconventional melting scenarios. The celebrated KTHNY theory of 2D melting
thereby predicts a continuous two-stage melting from
a crystalline to a hexatic
phase and, subsequently, from the hexatic to an isotropic liquid~\cite{KTHNY}.
The intermediate hexatic phase has short-range translational order
(i.e., it is fluid-like) but quasi-long-range orientational order,
characterized by a power-law decay of bond-angle pair correlations.
Other possibilities are a standard first-order 2D-melting transition
(as in Chui's theory~\cite{Chui}) or a stronger, i.e., first-order
hexatic-to-isotropic transition, as shown by Bernard and Krauth
to be the case for hard discs~\cite{Bernard}. In some cases, melting is
one-stage and first-order but the hexatic phase can be accessed via a
metastable route~\cite{Chen,Mejia}.

The KTHNY theory has been confirmed many times in simulation and experiment,
especially for particles with long-ranged interactions, though in a few cases
with some controversy about the order of the two transition
steps~\cite{Murray,Marcus,Zahn,Keim,Lin1,Han,Peng,Muto,Lin2,Lee,Qi,Clark}.
From a computational point of view, the assessment of the
order of 2D melting in practical cases can be hard, due to important
finite-size effects and slow relaxation to equilibrium. An extreme case
is hard discs where the exact nature of the melting transition could be
established only recently, by employing unprecedentedly large system sizes
of the order of a million particles in a box of fixed
volume~\cite{Bernard}. Generally speaking, working in the isothermal-isobaric
ensemble (rather than at constant volume) can ease the interpretation
of the simulation data, since one cannot incur in any of the
finite-size artifacts that harm constant-volume simulations.
Recently, we have provided unambiguous evidence of the occurrence of
continuous melting via a hexatic phase in the 2D Gaussian-core model
(GCM)~\cite{Prestipino1}. This system, taken as a prototype of the
phase behavior of ``empty'' polymers~\cite{Likos1}, has long been known
to exhibit reentrant melting (i.e., solid melting under isothermal
compression) as well as water-like anomalies in three
dimensions~\cite{Stillinger,Lang,Prestipino2,Mausbach}. In two dimensions,
the GCM melting becomes two-staged, with an extremely narrow hexatic
region whose properties comply with the predictions of the KTHNY theory.
In particular, besides a reentrant isotropic fluid, also a reentrant
hexatic phase exists. Recently, water-like anomalies have been reported
even in the one-dimensional GCM~\cite{Speranza}, where however no properly
defined thermodynamic transition occurs.

A class of systems which exhibit reentrant melting, solid polymorphism,
and other water-like anomalies is formed by particles interacting via
spherically-symmetric core-softened (CS) potentials~\cite{Buldyrev}.
For these systems, the strength of the repulsive component of the
interaction undergoes some sort of weakening over a range of interparticle
distances. The effects of particle-core softening on thermodynamic behavior
were originally investigated by Hemmer and Stell~\cite{Hemmer}, who were
interested in the possibility of multiple critical points and
isostructural solid-solid transitions in simple-fluid systems.
A few years later, Young and Alder~\cite{Young} showed that the phase diagram
of the hard-core plus square-shoulder repulsion exhibits an anomalous melting
line of the kind observed in Cs and Ce. Later, Debenedetti~\cite{Debenedetti}
showed that systems of particles interacting via continuous CS potentials
are capable of contracting when heated isobarically (a behavior which goes
under the name of ``density anomaly'').
In the last decade or so, there has been a renewal of interest in the phase
behavior of CS potentials, which has led to the discovery of many unusual
properties such as multiple reentrant-melting lines, polymorphism both in
the fluid and solid phases, stable cluster solids (for bounded interactions
only), and a plethora of thermodynamic and dynamic water-like
anomalies~\cite{Sadr,Jagla,Likos2,Kumar,Deoliveira,Gibson,Lomba,Fomin,Xu,Pizio,Pamies,Pauschenwein,Malescio1}.
A common feature of CS potentials is their ability to generate two distinct
length scales in the system, one related to the inner core and the other
associated with the milder component of the repulsion~\cite{Buldyrev}.
Due to this, CS fluids are characterized by two competing, expanded and
compact, local particle arrangements. Such an interplay has a disruptive
influence on crystal order, the fluid phase being thus recovered upon
compression.
This two-scale or two-fluid property mimics the behavior of the more complex
network-forming fluids, where the looser and denser local structures arise
from the incessant building and breaking of the dynamic network generated
by directional bonds. As a result, a paradigm has been established whereby
the existence of two competing local structures is essential for the
occurrence of anomalous behaviors in simple fluids.

Unusual phase behavior has also been found in 2D systems.
Scala and coworkers~\cite{Scala} carried out molecular-dynamics simulations
of the 2D square-shoulder plus
square-well (SSSW) potential finding water-like anomalies and identifying
two different solid phases: a triangular crystal (low-density phase) and a
square crystal (high-density phase).
However, the observation of a hysteresis loop in their simulations
suggested that the liquid-solid transition is actually first-order,
thus washing out the possibility of a continuous melting to a hexatic phase.
Wilding and Magee also studied the 2D SSSW potential and showed that the
thermodynamic
anomalies of the model, rather than stemming from a metastable liquid-liquid
critical point as previously surmised, were induced by the quasicontinuous
nature of the 2D freezing transition~\cite{Wilding}.
Almudallal and coworkers~\cite{Almudallal} contested this result, showing by
various free-energy techniques that all transitions in the 2D SSSW model are
in fact first-order.
Malescio and Pellicane studied a 2D system of particles interacting
through a potential consisting of an impenetrable hard core plus a square
shoulder. They found a variety of stripe phases whose formation was imputed
to the existence of two characteristic length scales~\cite{Malescio2}.
All such investigations suggest that the existence of two length scales
would be a prerequisite for observing water-like anomalies also in 2D.

In a series of recent papers the two-fluid
picture has been overthrown and a novel minimal scenario for the occurrence of
anomalies has been established~\cite{Saija,Prestipino3,Malescio3,Malescio4}.
More specifically,
it has been shown that a weakly-softened isotropic pair repulsion, with
a single characteristic length which becomes more loosely defined in a range
of pressures, is able to give rise to an unusual phase behavior.
Our purpose here is to verify whether anything similar occurs in 2D, where
in addition an interesting interplay with hexatic order may take place.
Far from being purely academic, the present study can be relevant for many
soft materials. For instance, one monolayer of $N$-isopropylacrylamide
(NIPA) microgel spheres confined between two glass cover slips is an ideal
system to study 2D melting because the effective interparticle potential
is short-ranged and repulsive, with a temperature-tunable volume
fraction~\cite{Dullens,Lin2,Han,Peng,Alsayed}.
Systems like this would be natural candidates for detecting, by video optical
microscopy or light scattering, the kind of phenomena that we are going to
illustrate below.

The plan of the paper is the following. After introducing our 2D model and
method in Section II, we sketch the model phase diagram in Section III,
highlighting the unconventional structure of the fluid phase which, similarly
to other instances of CS repulsion, shows a number of water-like anomalies
without any interplay between two characteristic length scales.
The melting mechanism is studied in more detail in Section IV,
where we show the existence of a hexatic phase.
Section V is finally devoted to concluding remarks.

\section{Model and method}
\setcounter{equation}{0}
\renewcommand{\theequation}{2.\arabic{equation}}

We consider a purely repulsive pair potential in two dimensions, modelled
through an exponential form which was first introduced, about four decades ago,
by Yoshida and Kamakura (YK)~\cite{Yoshida}:
\be
u(r)=\epsilon\exp\left\{a\left(1-\frac{r}{\sigma}\right)-
6\left(1-\frac{r}{\sigma}\right)^2\ln\left(\frac{r}{\sigma}\right)\right\}\,,
\label{2-1}
\ee
where $\epsilon$ and $\sigma$ set the energy and length scales, respectively,
and $a\ge 0$.
The YK potential behaves as $r^{-6}$ for small $r$, and falls off very rapidly
for large $r$. The smaller $a$ the higher the degree of softness of
$u(r)$, i.e., the flatter the repulsive ``shoulder'' around $r=\sigma$.
Technically speaking~\cite{Debenedetti}, $u(r)$ can be regarded as soft only
for $a\lesssim 2.3$, since only in this range there is an
interval of distances where the local virial function $-ru'(r)$ decreases
when the interparticle separation decreases.
However, it is over the wider range $a\lesssim 5.5$ that the phase diagram
is expected to show a reentrant-fluid region~\cite{Prestipino3,Prestipino4}.
We shall here focus on the case $a=3.3$ (Fig.\,1), which was already shown to
possess a rich anomalous phase behavior in three dimensions which cannot
simply be explained by the existence of two distinct nearest-neighbor (NN)
distances in the dense fluid~\cite{Prestipino3}.

We now briefly describe the methodology of the present investigation.
After identifying the relevant crystal phases by means of total-energy
calculations at zero temperature
(of the type illustrated in, e.g., Ref.\,\cite{Prestipino5}),
we explore the phase diagram by Monte Carlo (MC) simulation in the $NPT$
ensemble. As is common practice, we adopt periodic boundary conditions and
employ cell linked lists in order to speed up the simulation.
While systems of about 1000 particles are suitable for investigating
bulk properties and for determining the approximate location of
phase boundaries, we consider larger systems of about 6000 particles
for the search and characterization of the hexatic phase at selected
pressures. For the same $P$ values, we check the order of the melting
transition independently through thermodynamic integration combined
with exact free-energy calculations.
For the fluid phase, a dilute gas is used as a reference state,
whose chemical potential is calculated by the Widom method, whereas a
low-temperature crystal is chosen as the starting point of the simulation in
the solid region of the phase diagram. At this initial state, the excess
Helmholtz free energy of the system is computed by the Einstein-crystal
method~\cite{Frenkel}.

\section{Fluid structure and water-like anomalies}

For $a=3.3$, the repulsive shoulder at $\sigma$, which is still visible
in the plot of $u(r)$ for $a\simeq 2$, has by then completely faded out,
surviving only in the form of a modest bump in the otherwise
monotonously-decreasing profile of the virial function (see Fig.\,1).
Yet, this almost structureless potential shows three distinct stable crystal
arrangements at $T=0$ in two dimensions, as it follows from the
calculation of the chemical potential $\mu$ as a function of the pressure $P$
for all five Bravais lattices and for the honeycomb lattice.
Upon increasing $P$, the sequence of phases is triangular-square-triangular,
the square crystal being thermodynamically stable in the pressure range
$2.27$-$4.36$, in units of $\epsilon/\sigma^3$ (from now on, pressure and
temperature will be given in reduced units).

Assuming that no other crystal phases come into play at nonzero temperatures,
we first sketched the overall phase diagram by the heat-until-it-melts
(HUIM) method. In practice, for selected $P$ values we run a chain of MC
simulations of the solid stable at the given $P$, for increasing $T$ values
at regular intervals of 0.0005, until we observe a clear
jump in both the particle-number density $\rho=N/V$ and total energy per
particle $E/N$, which signal the melting of the solid. The character of this
transition will be addressed in the next section. The HUIM method simply
overlooks the possibility of solid overheating, but this is usually a
reasonable
approximation if one is not interested in very precise estimates of the
melting temperature $T_m$ ({\it a posteriori}, the typical
error implied by the HUIM method in determining the melting point of the
present system was about 10\%, and smaller for the square crystal).
The outcome of this analysis
is reported in Fig.\,2 where the most revealing feature is the reentrance
of the fluid phase as the system, already settled in the low-density
triangular solid, is further compressed at not too low temperatures.
Fluid reentrance occurs even twice in a smaller range of temperatures above
0.020.

As we already said in the Introduction, until very recently the conventional
wisdom on the origin of the reentrant-melting phenomenon in CS systems
with an unbounded interparticle repulsion rested on a competition,
with destabilizing effects on crystal ordering, between two ways of
arranging particles close to each other, which gives rise to either an
expanded or a more compact structure in the dense fluid. This may only occur
provided the pressure is strong enough as to bring neighboring particles at
distances close to $\sigma$. Clearly, this explanation cannot work for
a case like the present one, where it is hard to maintain that there are
two different length scales in the potential. Rather, we may view the
question from a solid-state perspective and argue that a succession of
reentrant-melting lines in the phase diagram is simply the outcome of
the existence of multiple solid phases at low temperature. On increasing
pressure within the range of stability of any of such solids, the crystal
strength first grows up to a maximum (and in parallel also $T_m$),
but then it progressively reduces on approaching the boundary
of the next stable solid at higher pressure. After all, this is exactly
the rationale behind the melting criterion discussed in
Ref.\,\cite{Prestipino4},
which in fact is very effective for soft repulsive interactions.

We found confirmation that the present model shows only one repulsive
length scale by computing the fluid radial distribution function (RDF)
along the isotherm at $T=0.1$, i.e., just above the maximum $T_m$
of the low-density triangular solid (see Fig.\,3). Looking
at the position of the main RDF peak, we see a systematic shift to lower
and lower distances upon compression, with a slight widening of the peak
around $P\simeq 2$, i.e., next to the first reentrant-melting line.
We interpret this as the evidence of a unique NN characteristic distance
in the system, which becomes more loosely defined across the region of
reentrant melting. This should be contrasted with what occurs for a more
conventional CS repulsion, where the first RDF maximum is twin-peaked,
with two definite NN distances that take turns at providing the absolute
maximum for the RDF.

Aside from the existence of one repulsive length scale rather than two,
many water-like anomalies are also found in the present
system, starting with a line of $\rho$ maxima in the
fluid close to the first reentrant-melting line (see Fig.\,2).
This is a typical occurrence in systems with CS potentials,
where however the region of volumetric anomaly is usually more extended.
We checked that the line of the density anomaly ceases to exist just before
plunging into the solid region. Another type of anomaly is the
so-called structural anomaly, that is a non-monotonous pressure behavior
of the amount of ``translational order'' in the fluid, as measured through
the value of minus the pair entropy per particle ($-s_2$)~\cite{Prestipino6}
(see Fig.\,2 and the left top panel of Fig.\,4).
Rather than monotonously increasing with pressure at constant temperature,
the degree of spatial order reduces in the reentrant-fluid region, as a
result of a looser definition of the NN distance.
A non-sharp average separation between neighboring particles
acts as a perturbing factor for the local order, bringing about a slight
decrease of $-s_2$ with pressure.
In the same range of pressures where $-s_2$ decreases,
the self-diffusion coefficient $D$, which we
measure by a series of $TVN$ molecular-dynamics runs, gets enhanced with
pressure (Fig.\,2 and right top panel of Fig.\,4). Both lines of structural
and diffusional anomalies appear to sprout out of the point of maximum $T_m$
for the low-density triangular solid and roughly terminate at the point of
maximum $T_m$ for the square solid. This is again similar
to other CS systems, also for what concerns the crossing of the anomaly lines
(see e.g. Fig.\,11 of \cite{Jabes}). Finally, we checked for just one pressure
value ($P=2$) the existence of a temperature minimum in both the isobaric
specific heat $C_P$ and isothermal compressibility $K_T$ (lower panels
of Fig.\,4), similarly to what found for water at ambient pressure.
Both minima fall in the high-temperature fluid region,
much far from the maximum-density point for the same pressure.

\section{Hexatic behavior}

The existence in our model of two solids with different crystal symmetry,
i.e., triangular and square, gives the opportunity to investigate the
intriguing
possibility of two distinct intermediate phases (hexatic and tetratic)
between the solid and the normal fluid, with the further bonus of a
yet-to-be-observed
hexatic-tetratic transition near the confluence point between the two solids
and the fluid.

In order to clarify the melting scenario for a given $P$,
we run our computer code along two simulation paths, one starting
from the solid phase at $T=0.005$ and the other from a high-temperature
fluid state. We advance in steps of $\Delta T=0.005$, equilibrating the
system for long before generating an equilibrium trajectory of $M$ sweeps
(one sweep corresponding to $N$ trial MC moves), with $M$ ranging between
$5\times 10^5$ and $3\times 10^6$, depending on how far we are from melting.
Once a guess of the transition point is made, we restart the simulation
slightly ahead of it with a smaller $\Delta T$ and/or a larger $M$
in order to better discriminate between first-order and continuous melting.
Besides $\rho$ and $E/N$, we measure two order parameters (OP),
$\psi_T$ and $\psi_O$, which are sensitive to the overall translational
and orientational triangular/square order, respectively. The precise
definition of both quantities has been given in Ref.\,\cite{Prestipino1},
with obvious modifications for the square-lattice case. Moreover, we keep
track of the OP susceptibilities $\chi_T$ and $\chi_O$, defined as $N$
times the variance of the respective OP estimator. Finally, we calculate
two orientational correlation functions (OCF)~\cite{Prestipino1}, $h_6(r)$
and $h_4(r)$, which inform on the typical size of a space region in which
NN-bond angles are strongly correlated. The KTHNY theory predicts an
algebraic $r^{-\eta(T)}$ large-distance decay of the OCF in the hexatic
phase, at variance with what occurs in an isotropic fluid where the decay
is much faster, i.e., exponential. According to the same theory, $\eta$
equals $1/4$ at the transition point between hexatic and isotropic fluid.

In Figs.\,5-8, we report $\rho$ and $E/N$ for two different system sizes
and various simulation protocols, as a function of $T$ for
$P=0.5,2,3$, and 5. We clearly see that, while melting is
continuous for $P=0.5,2$, and 5, it is
certainly first-order for $P=3$ (and 4, data not shown), as evidenced by the
hysteresis loops. Based on our experience with the GCM~\cite{Prestipino1},
we can conjecture that there is a narrow region of hexatic phase for
$P=0.5,2$, and 5
(to be confirmed later by the analysis of OPs and OCF), whereas no
definite conclusion can be reached at this point for $P=3$, where a
tetratic phase can in principle exist even in presence of a first-order
transition. A special remark is due for $P=5$, where,
in contrast with what occurs for smaller pressures, the crystal energy
decreases upon heating, which means that the increase in kinetic energy
is more than compensated for by the loss in potential energy, whose high
rate of decrease is due to a NN distance lying in the harsh part of the
potential core.

For three pressures ($0.5,2$, and 3),
we plot the OPs and related
susceptibilities in Figs.\,9-11. For the first two pressures, we see that
$\psi_T$ vanishes at a slightly smaller temperature than $\psi_O\equiv\psi_6$,
which implies that the hexatic phase is confined to a narrow $T$ interval
not wider than 0.0005 for $P=0.5$ (0.0015 for $P=2$), as also witnessed
by the maxima of the two susceptibilities occurring at slightly different
$T$ values. These temperature intervals compare well with the $T$ range
of the bridging region between the solid and fluid branches in Figs.\,5
and 6. We thus confirm the same phenomenology of the GCM, namely that the
width of the hexatic region increases with pressure.
For $P=5$, the findings are similar to $P=2$, with roughly
the same width of the hexatic region and comparable levels of orientational
and translational order in the hot solid (data not shown).
Going to $P=3$, the picture is quite different since the two OPs now apparently
vanish at the same temperature; at that point, the orientational
susceptibility shows a spike (rather than a critical peak)
which is usually associated with a first-order transition.
All evidence suggests that standard first-order melting is a plausible
explanation for $P=3$, and the same conclusion can be made for
$P=4$ (data not shown).

A more direct evidence of the existence of a bond-angle ordered fluid
(or a clue to its absence) comes from the large-distance behavior of
the OCF. We plot this function in Figs.\,12-14
for various temperatures across the relevant
region, for $P=0.5,2$, and 3, respectively. It appears that, for $P=0.5$
and 2, $h_6(r)$ decays algebraically in a $T$ region of limited extent,
roughly corresponding to the bridging region in Figs.\,5 and 6.
Moreover, the decay exponent in the hexatic region is smaller than
$1/4$, becoming larger only on transforming to the isotropic fluid,
and the same is found for $P=5$ (data not shown).
On the contrary, for $P=3$ and 4, $h_4(r)$ switches directly from no decay
at all to an exponential damping, showing that there is no tetratic phase
in our model.

The obvious question now arises as to why, at variance with the hexatic one,
the tetratic phase is not stable in our model. This question is difficult
to answer since
it involves the consideration of the delicate equilibrium between the energy
and entropy of the two competing phases, in this case the tetratic phase and
the isotropic fluid phase. A possible hint could be the level of orientational
order that is found in the two types of solid slightly before melting.
If we look at Figs.\,9-11, we see that the value of $\psi_O\equiv\psi_4$
for $P=3$ is less than a
half of $\psi_6$ (note that the same holds for $P=4$), and
this would explain why, in the square-crystal case, long-range orientational
order does not survive (even in the weakened form typical of a tetratic
phase) the loss of quasi-long-range positional order determined by melting.

Finally, we checked by an indipendent route the order of the melting
transition for $P=0.5,2,3$, and 4. As anticipated, for each pressure we
carried out MC simulations along two different paths, one beginning
from a cold solid and the other from a low-density fluid.
Using thermodynamic integration in combination with exact free-energy
calculations at the initial points of the paths, we were able to obtain
the system chemical potential $\mu$ along the solid and fluid branches
as a function of $T$. For $P=0.5$ and 2, we did not find any crossing
of the two $\mu(T)$ curves, thus confirming a continous melting transition.
Notwithstanding the care we put in keeping under control any source of
statistical error, we nevertheless found a small discrepancy (about
$0.0005\epsilon$, practically constant in a 0.01 wide $T$ range
across the melting transition) between the $\mu$ levels of the solid
and the fluid, which is presumably due to the not-so-small $\Delta T$
employed along the paths far from melting. On the contrary, for $P=3$
and 4, we observed a clear crossing between the two $\mu(T)$ curves,
respectively at $T=0.0269$ and $T=0.0176$, which is consistent with
the location of the density and energy jumps (for $P=3$, see Fig.\,7),
and suggestive of a first-order transition.

Summing up, we collected multiple evidence of a continuous
melting via a hexatic phase for the triangular crystals,
while the melting transition is certainly discontinuous and ``standard''
for the square crystal. Nothwithstanding the ``small''
sizes of the investigated samples, we think that our conclusions are robust
since they result from many independent indicators of the phase-transition
order.

\section{Conclusions}

We have analyzed the phase behavior of purely-repulsive 2D particles
with a weakly-softened core. The nature of this repulsion is such as to
determine a characteristic nearest-neighbor distance in the fluid phase
whose statistical precision, expressed by the width of the main peak of
the radial distribution function, shows a non-monotonous behavior with
pressure at not too high temperatures. Notwithstanding the fact that the
two-fluid paradigm of core-softened (CS) potentials does not apply here,
we anyway observe the same phenomenology as in conventional CS systems,
with solid polymorphism, multiple reentrant-melting lines, and many other
water-like anomalies.
While this is similar to the 3D case~\cite{Prestipino3}, the melting
transition of the present system is different, since it is continuous
and occurs via an (even reentrant) hexatic phase for the triangular
solids while being standard first-order for the
intermediate square solid (i.e., no parallel tetratic
phase exists). The hexatic behavior appears to be
consistent with the KTHNY theory,
as witnessed by the value of the decay exponent of the orientational
correlation function at the hexatic-to-isotropic fluid transition temperature.

Our findings could be relevant for many real soft-matter systems.
Already nowadays, monodisperse colloidal suspensions can be engineered
in such a way as to exhibit a temperature-modulated repulsion with some
amount of softness. Probably, the day is not far when by appropriate
functionalization of the particle surface a colloidal system will be shown
to exhibit the kind of phase-transition phenomena that we have illustrated
here for a rather specific model potential.

\newpage
%
%
\begin{figure}
\includegraphics[width=9cm]{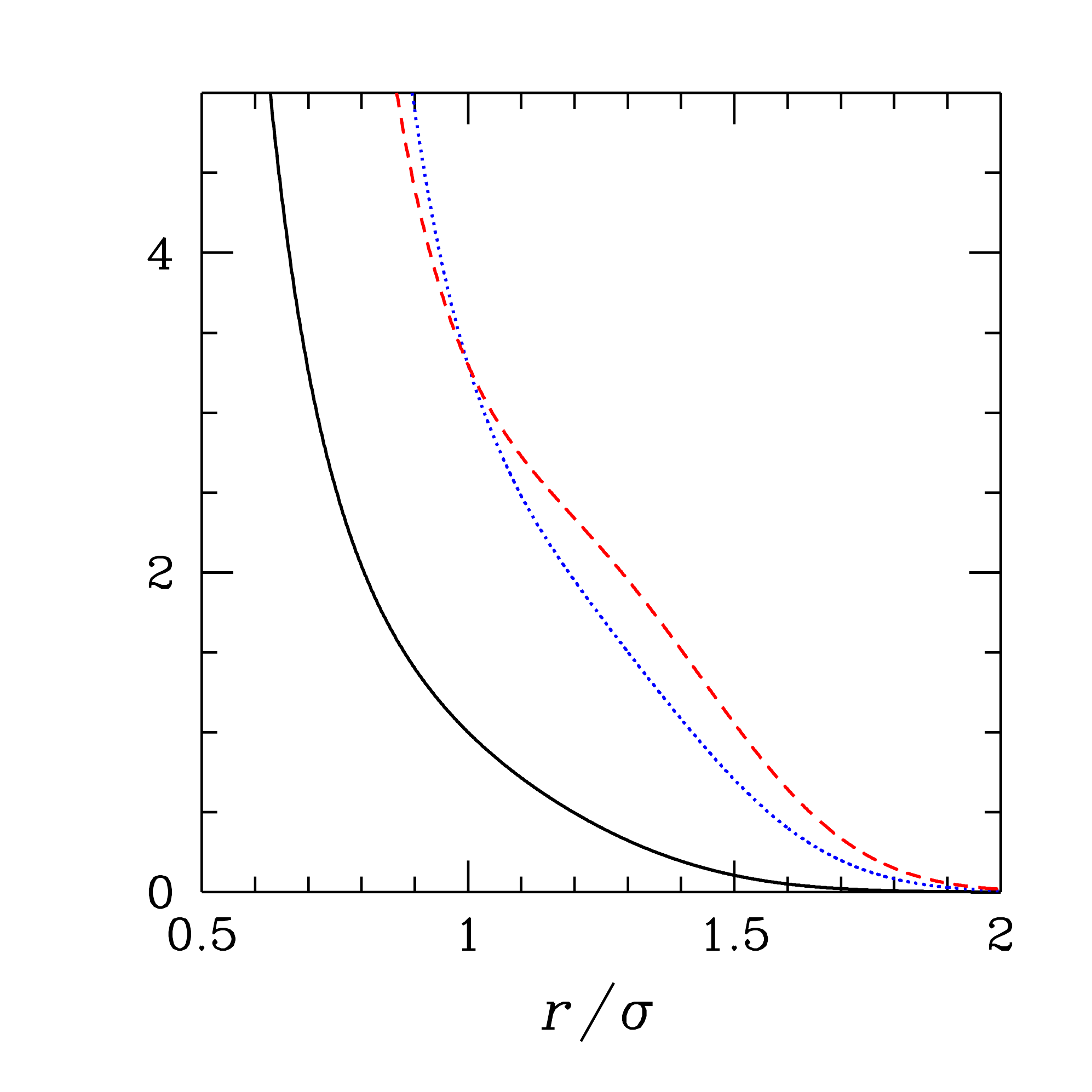}
\caption{(Color online). YK potential with $a=3.3$ (Eq.\,(\ref{2-1}), black
solid line), its negative first-order derivative $-u'(r)$ (dotted blue line),
and the virial function $-ru'(r)$ (dashed red line).}
\label{fig1}
\end{figure}

%
%
\begin{figure}
\includegraphics[width=9cm]{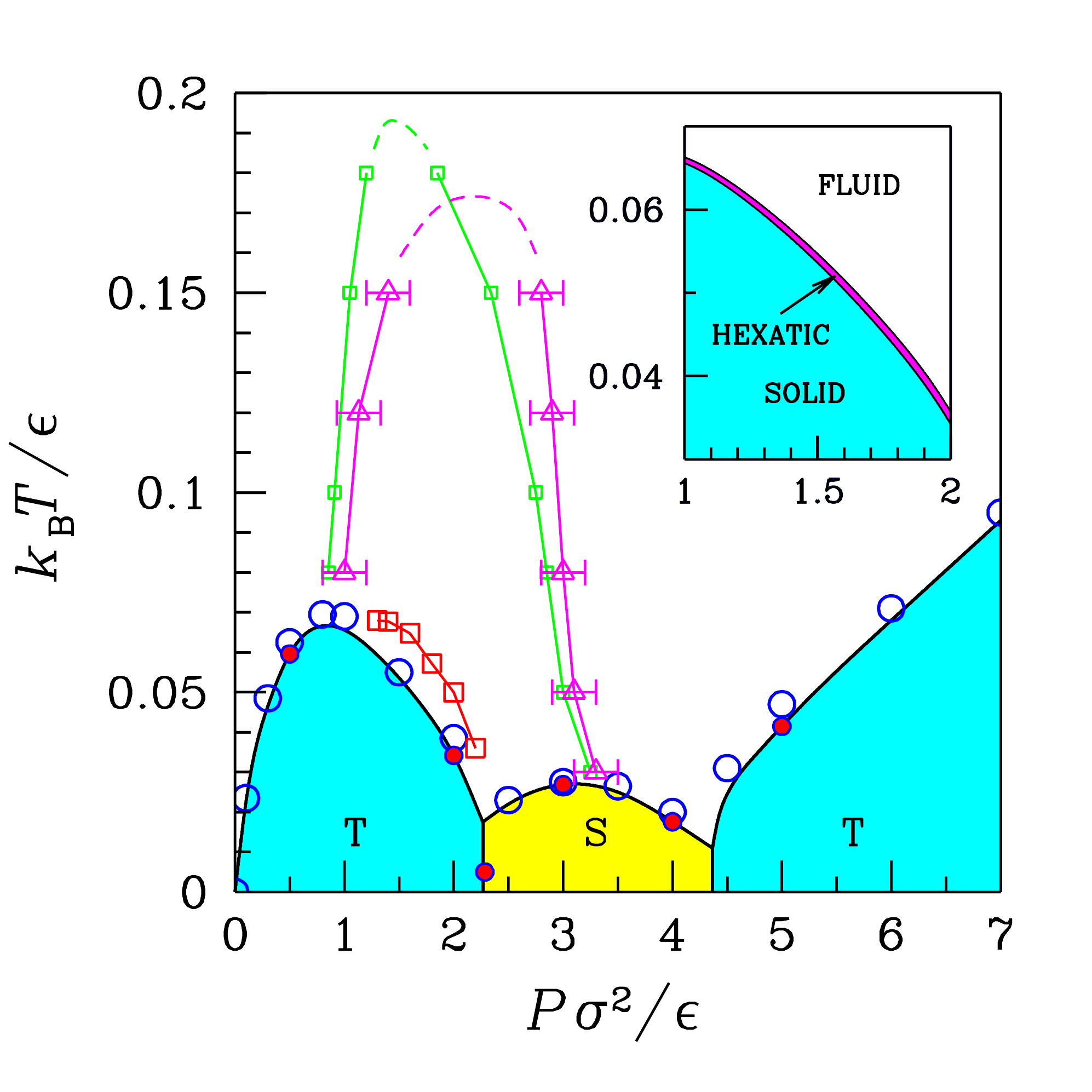}
\caption{(Color online). Phase diagram of the YK potential with $a=3.3$.
Two kind of melting points are shown: those determined by the HUIM method
(blue open dots) and others (red-filled blue dots) marking either the onset
of the hexatic phase ($P=0.5,2$, and 5) or the crossing in
temperature of solid and liquid chemical potentials ($P=3$ and 4), as
discussed in Section IV. Statistical errors are smaller
than the symbols size. Another red dot is placed where the chemical
potential of the low-density triangular (T) solid takes over that of the
square (S) solid at $T=0.005$. The black solid lines are
schematic transition lines. The hexatic-isotropic fluid transition
lines are not shown.
The extent of the low-pressure hexatic region can be
appreciated in the inset, which shows a magnification of the $P$ interval
from 1 to 2. Of similar width is the high-pressure hexatic region near $P=5$.
Further open symbols joined by straight lines mark the boundaries of
anomaly regions: isothermal $-s_2$ maxima
and minima (left and right green squares);
isothermal $D$ minima and maxima (left and right magenta triangles);
isobaric $\rho$ maxima (red squares).
We checked that the structural and diffusional anomalies are no longer present
for $T=0.2$ and $T=0.18$, respectively; hence, we draw the dashed lines to
mean that the two branches of each anomaly are actually sewed in from the
top.}
\label{fig2}
\end{figure}

%
%
\begin{figure}
\includegraphics[width=9cm]{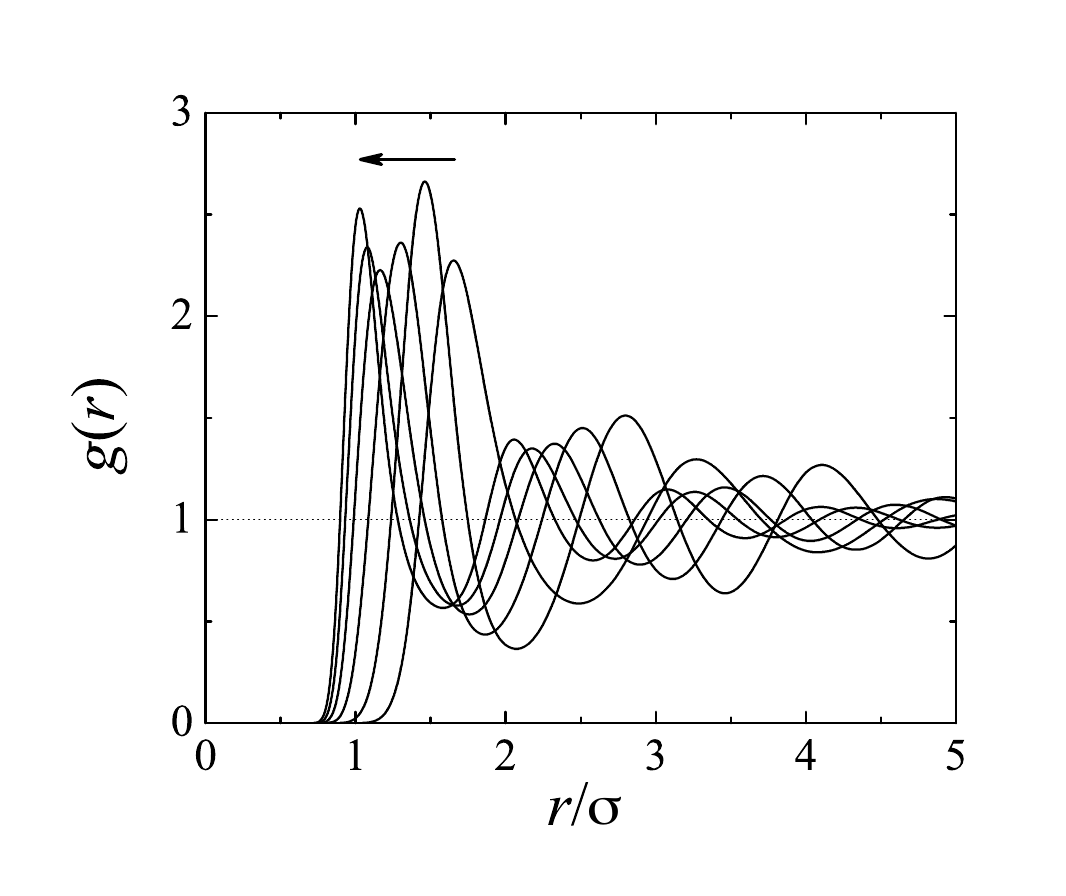}
\caption{YK potential with $a=3.3$: radial distribution function $g(r)$
for various pressures and for $T=0.08$. Successive lines correspond
to $P=0.2,0.8,1.6,2.4,3.2,4$. The arrow marks the direction along which
the pressure increases.}
\label{fig3}
\end{figure}

%
%
\begin{figure}
\includegraphics[width=9cm]{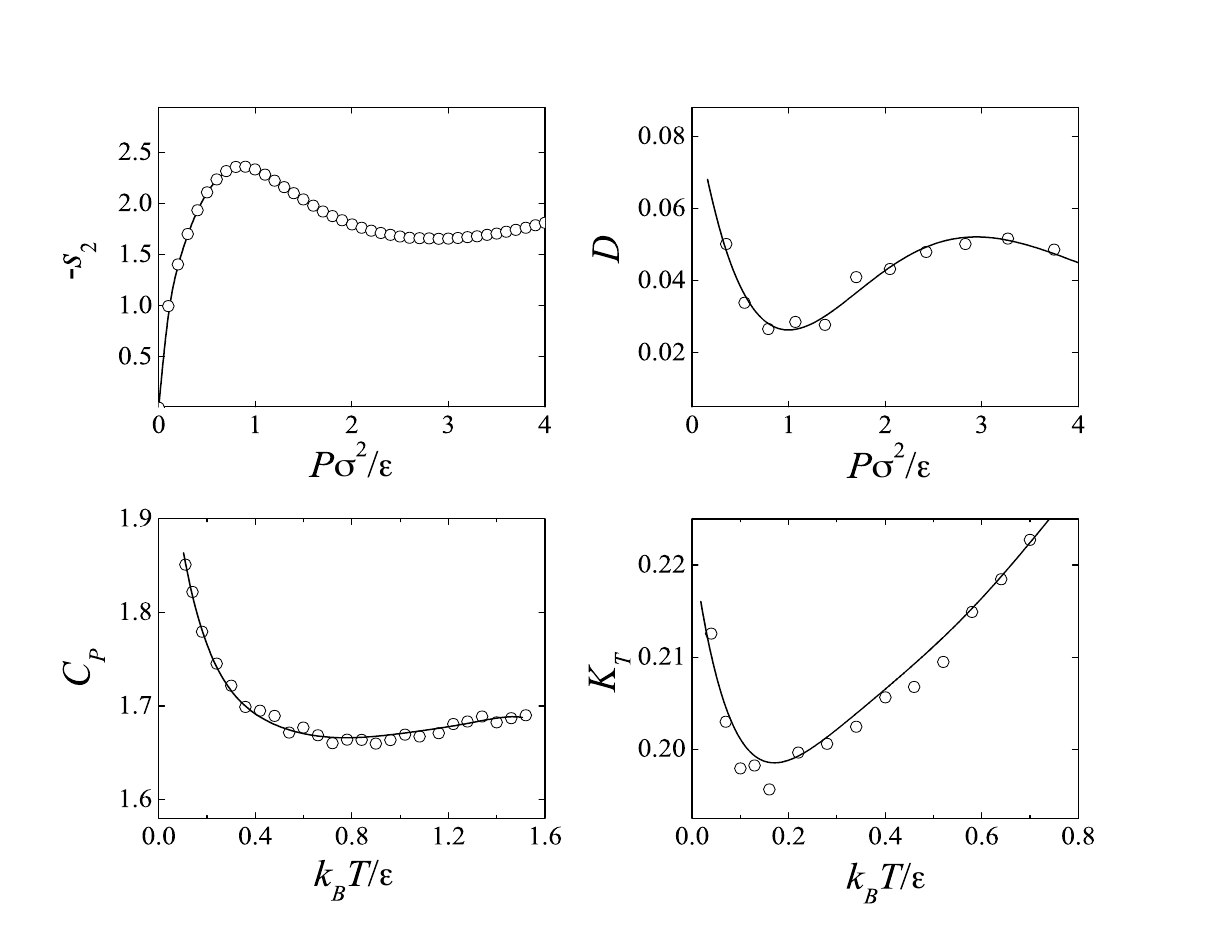}
\caption{Anomalous properties of the YK potential with $a=3.3$.
Left top panel: translational-order parameter $-s2$ (in units of $k_B$)
as a function of pressure for $T=0.08$. Right top panel: self-diffusion
coefficient $D$ (units of $\sigma(\epsilon/m)^{1/2}$, where $m$ is the
particle mass) for $T=0.08$. Left bottom panel: isobaric specific heat
$C_P$ (units of $k_B$) as a function of temperature for $P=2$.
Right bottom panel: isothermal compressibility $K_T$ (units of
$\sigma^2/\epsilon$) for $P=2$. A clear minimum is seen
in the plot of both response functions.}
\label{fig4}
\end{figure}

%
%
\begin{figure}
\includegraphics[width=9cm]{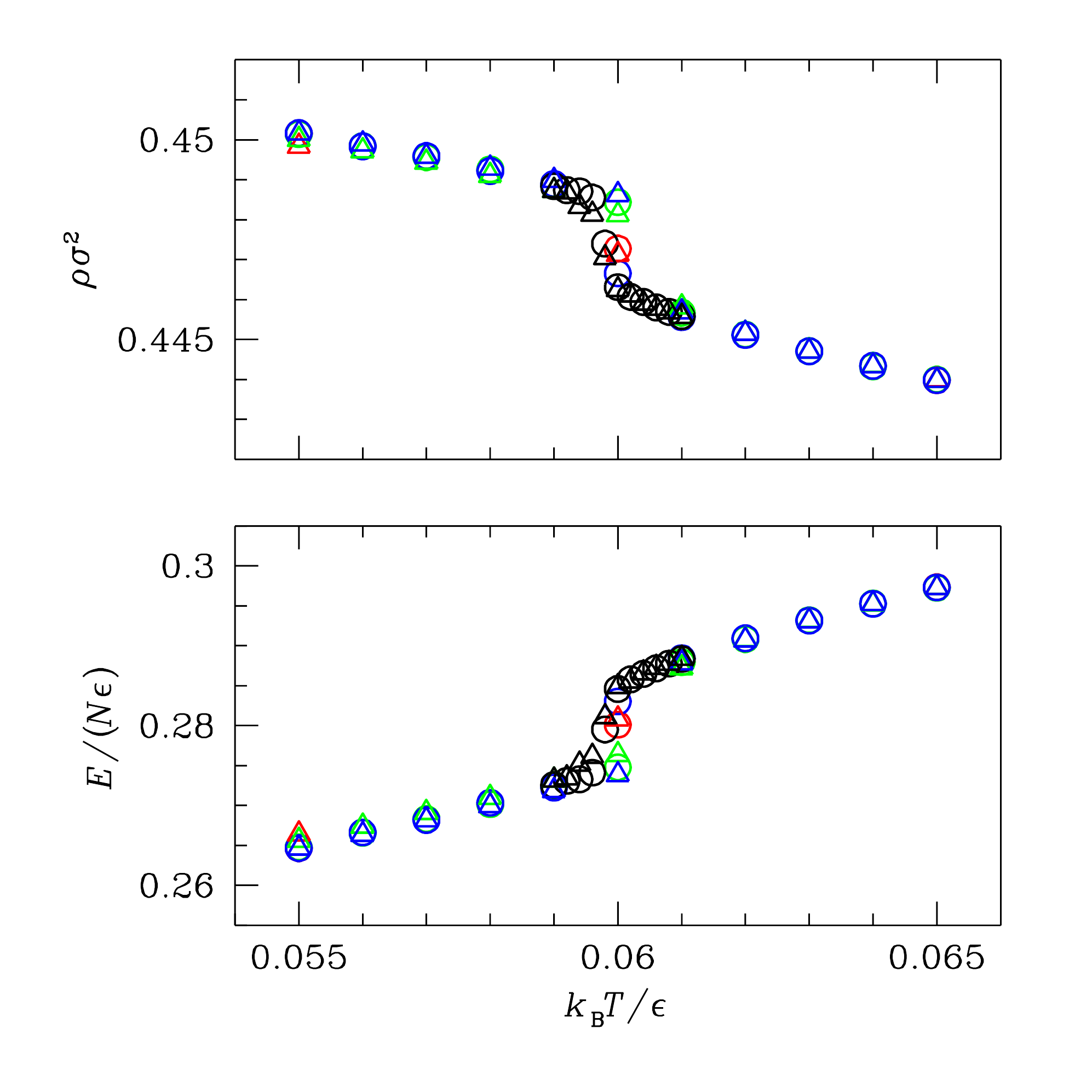}
\caption{(Color online). YK potential with $a=3.3$ for $P=0.5$: particle-number
density (top) and total energy per particle (bottom) for two different sizes
($N=2688$, red, green, and blue; $N=6048$, black). Different colors denote
different simulation protocols (red: $\Delta T=0.005$ and $M=5\times 10^5$;
green: $\Delta T=0.001$ and $M=5\times 10^5$; blue: $\Delta T=0.001$ and
$M=2\times 10^6$; black: $\Delta T=0.0002$ and $M=3\times 10^6$).
Open dots and triangles refer to a heating path and a cooling path,
respectively.}
\label{fig5}
\end{figure}

%
%
\begin{figure}
\includegraphics[width=9cm]{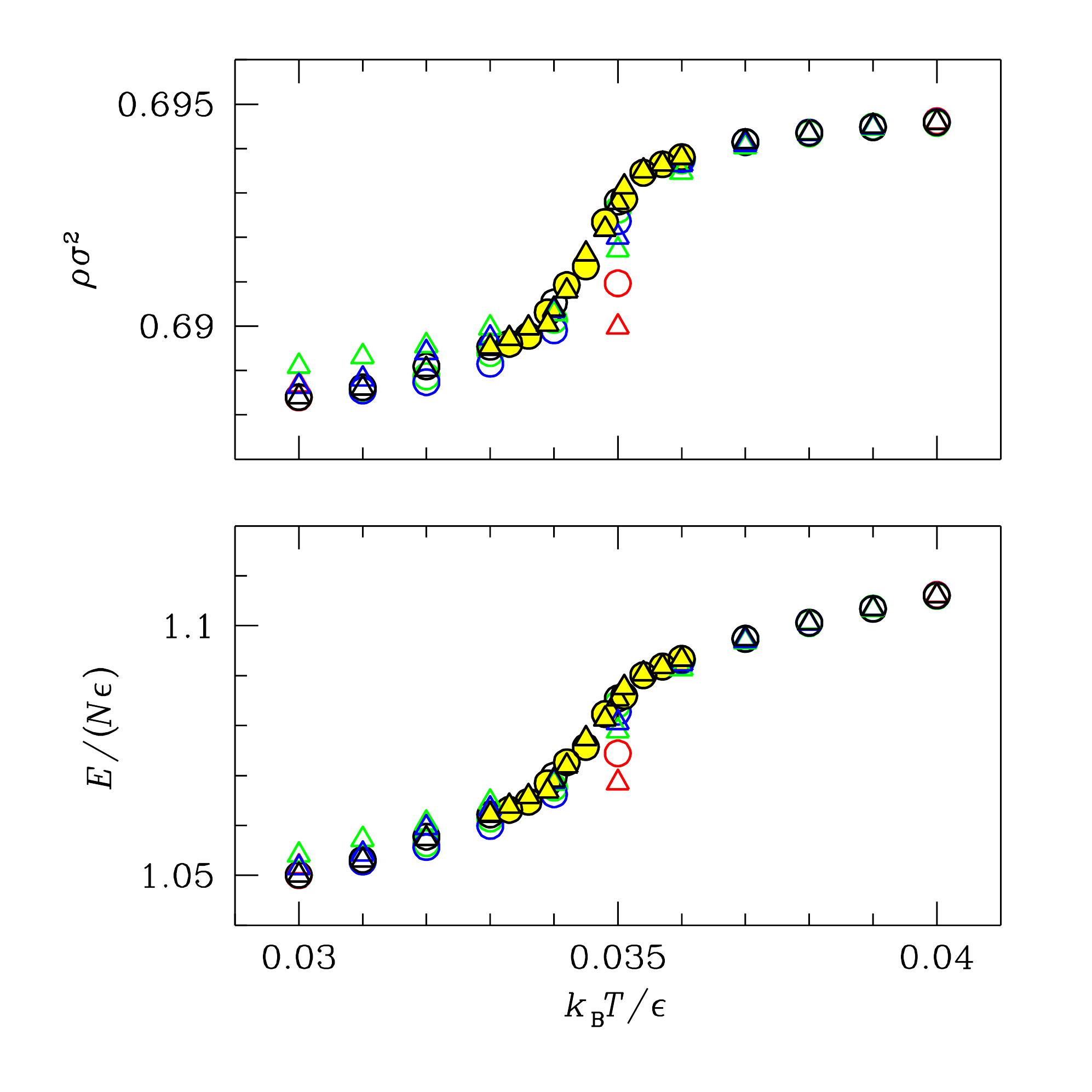}
\caption{(Color online). YK potential with $a=3.3$ for $P=2$: particle-number
density (top) and total energy per particle (bottom) for two different sizes
($N=2688$, red, green, and blue; $N=6048$, black and yellow-filled black).
Different colors denote different simulation protocols
(red: $\Delta T=0.005$ and $M=5\times 10^5$; green: $\Delta T=0.001$ and
$M=5\times 10^5$; blue: $\Delta T=0.001$ and $M=2\times 10^6$;
black: $\Delta T=0.001$ and $M=3\times 10^6$;
yellow-filled black: $\Delta T=0.0003$ and $M=3\times 10^6$).
Open dots and triangles refer to a heating path and a cooling path,
respectively.}
\label{fig6}
\end{figure}

%
%
\begin{figure}
\includegraphics[width=9cm]{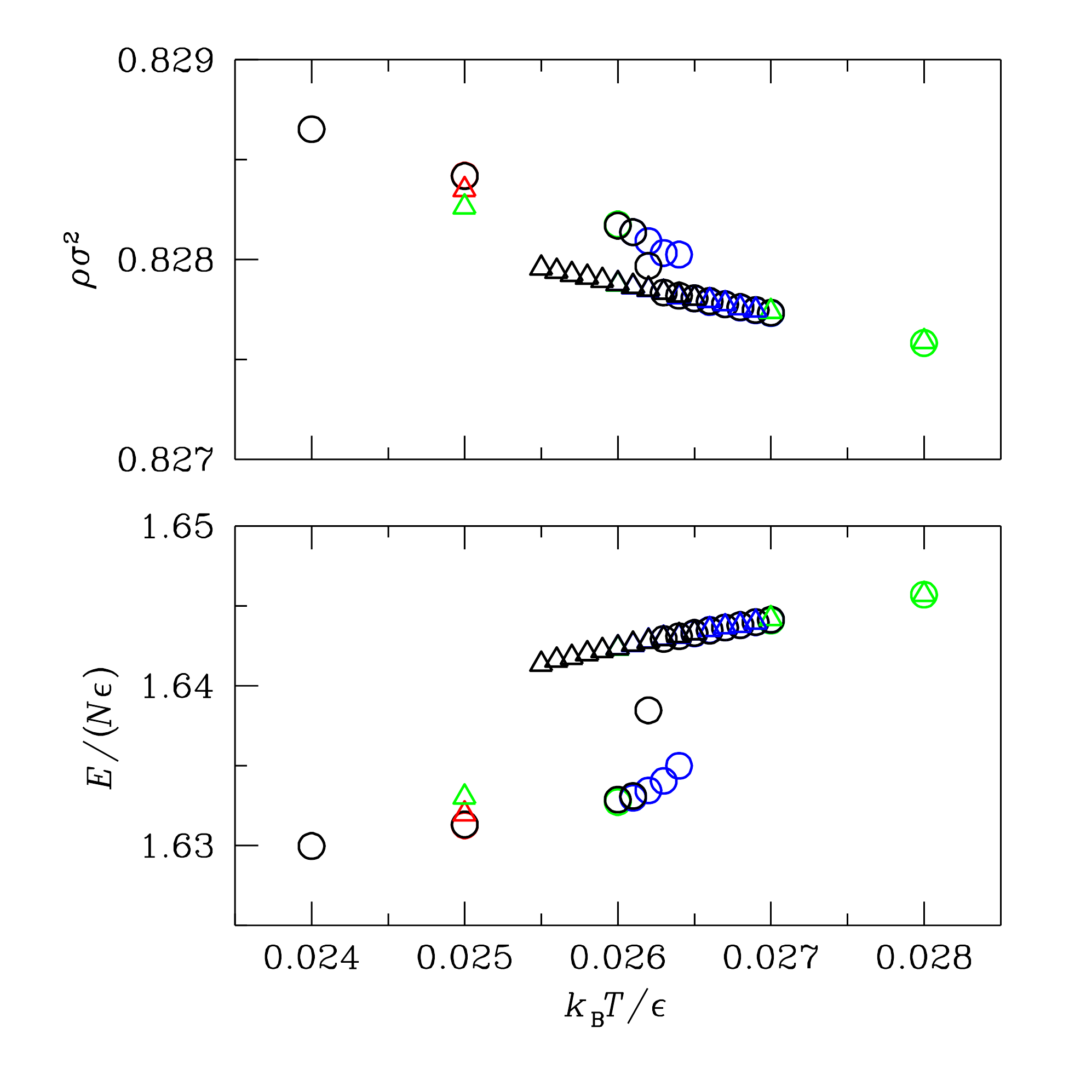}
\caption{(Color online). YK potential with $a=3.3$ for $P=3$: particle-number
density (top) and total energy per particle (bottom) for two different sizes
($N=2704$, red, green, and blue; $N=6084$, black). Different colors denote
different simulation protocols (red: $\Delta T=0.005$ and $M=5\times 10^5$;
green: $\Delta T=0.001$ and $M=5\times 10^5$; blue: $\Delta T=0.001$ and
$M=2\times 10^6$; black: $\Delta T=0.0001$ and $M=3\times 10^6$).
Open dots and triangles refer to a heating path and a cooling path,
respectively.}
\label{fig7}
\end{figure}

%
%
\begin{figure}
\includegraphics[width=9cm]{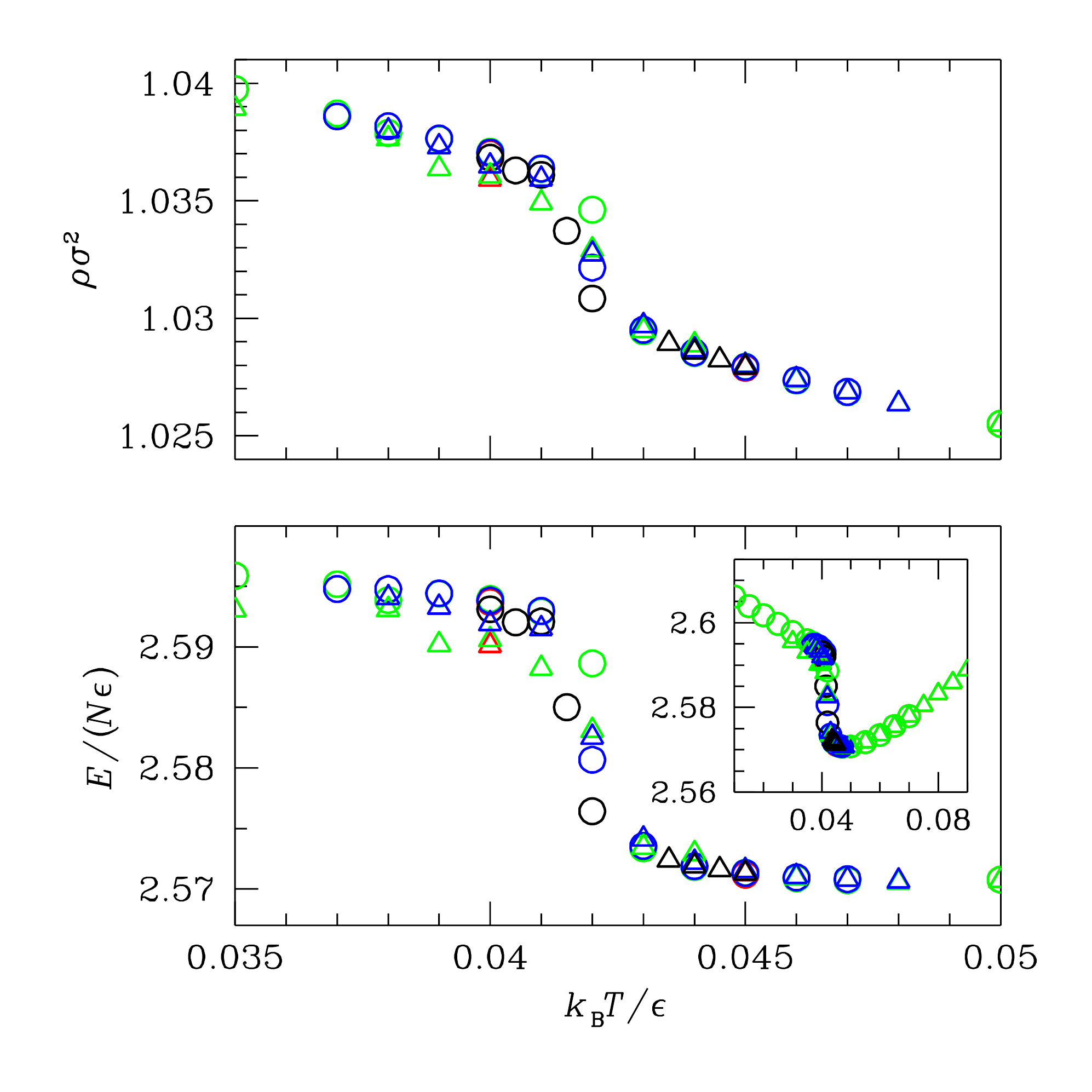}
\caption{
(Color online). YK potential with $a=3.3$ for $P=5$: particle-number
density (top) and total energy per particle (bottom) for two different sizes
($N=2688$, red, green, and blue; $N=6048$, black).
Different colors denote different simulation protocols
(red: $\Delta T=0.005$ and $M=5\times 10^5$; green: $\Delta T=0.001$ and
$M=5\times 10^5$; blue: $\Delta T=0.001$ and $M=2\times 10^6$;
black: $\Delta T=0.0005$ and $M=2\times 10^6$). Open dots and triangles
refer to a heating path and a cooling path, respectively. Inset, total
energy per particle on a wider pressure range.}
\label{fig8}
\end{figure}

%
%
\begin{figure}
\includegraphics[width=9cm]{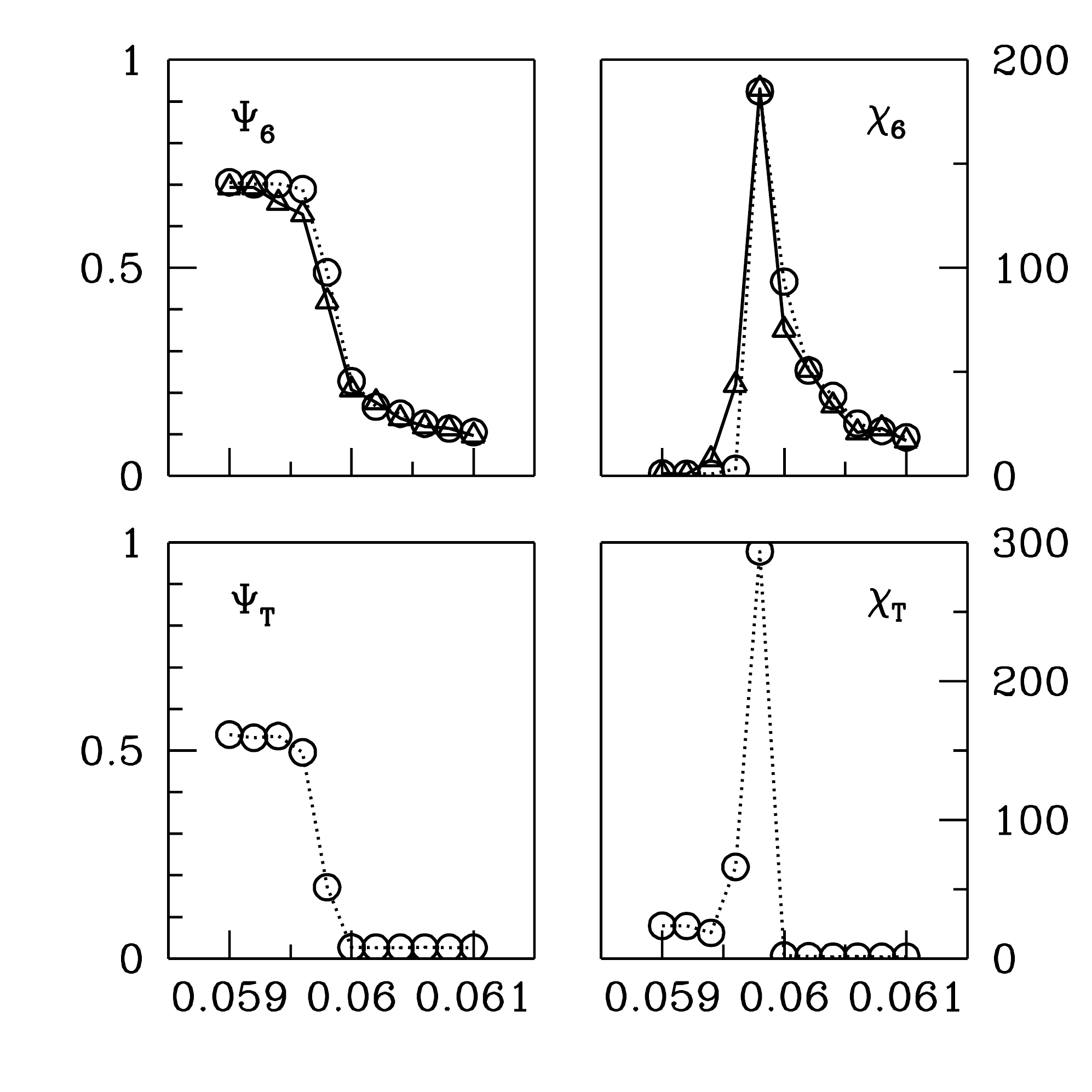}
\caption{YK potential with $a=3.3$ for $P=0.5$: order parameters and
susceptibilities in the $T$ range across the melting transition.
Upper panels: the orientational order parameter $\psi_6$ and its
susceptibility $\chi_6$. Dots and triangles mark data obtained by
heating and by cooling, respectively. Lower panels: the translational
order parameter $\psi_T$ and its susceptibility $\chi_T$ on heating.
All data from different protocols and sizes are reported, always
preferring the most accurate estimate when more than one is available.}
\label{fig9}
\end{figure}

%
%
\begin{figure}
\includegraphics[width=9cm]{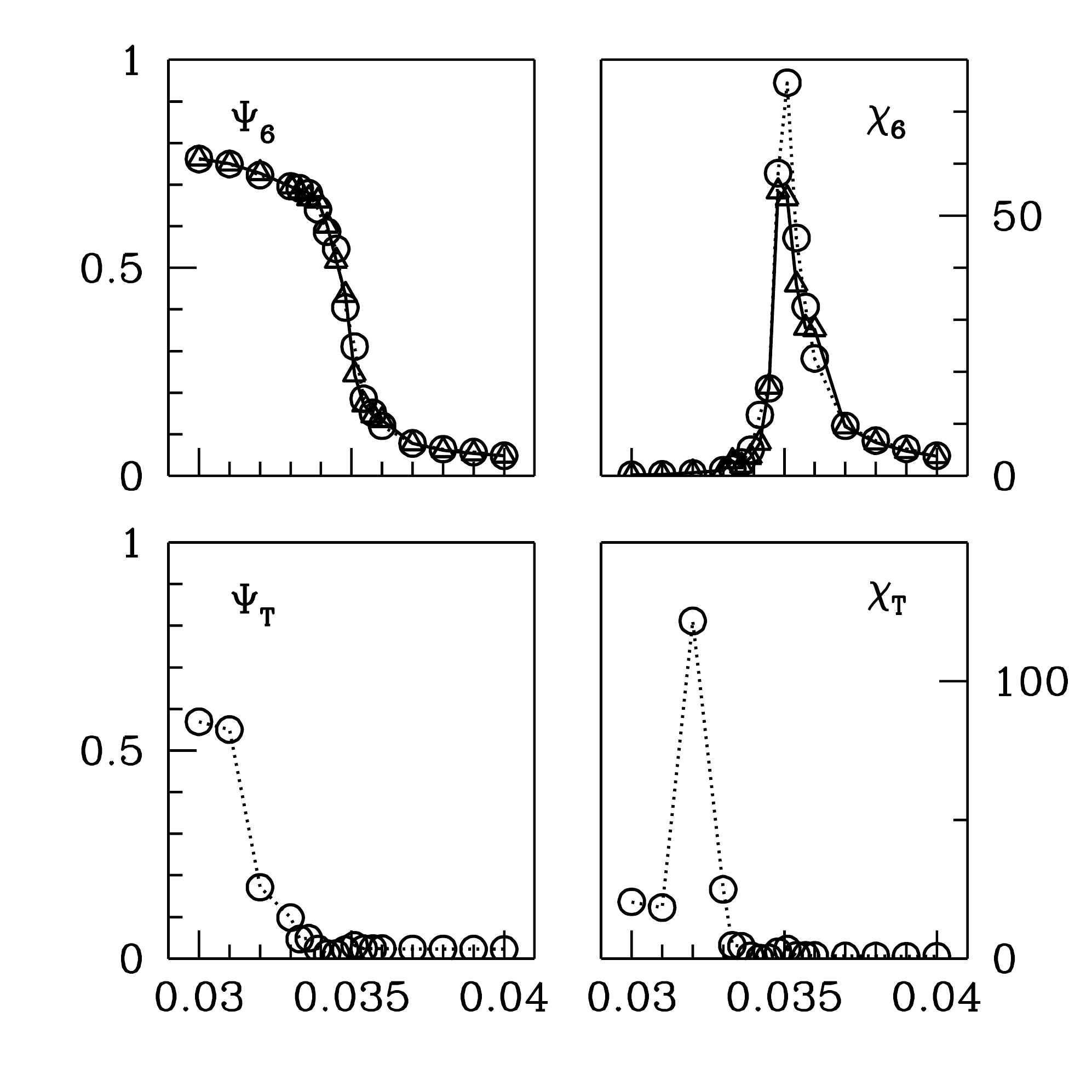}
\caption{YK potential with $a=3.3$ for $P=2$: order parameters and
susceptibilities in the $T$ range across the melting transition.
See the caption of Fig.\,9 for notation.}
\label{fig10}
\end{figure}

%
%
\begin{figure}
\includegraphics[width=9cm]{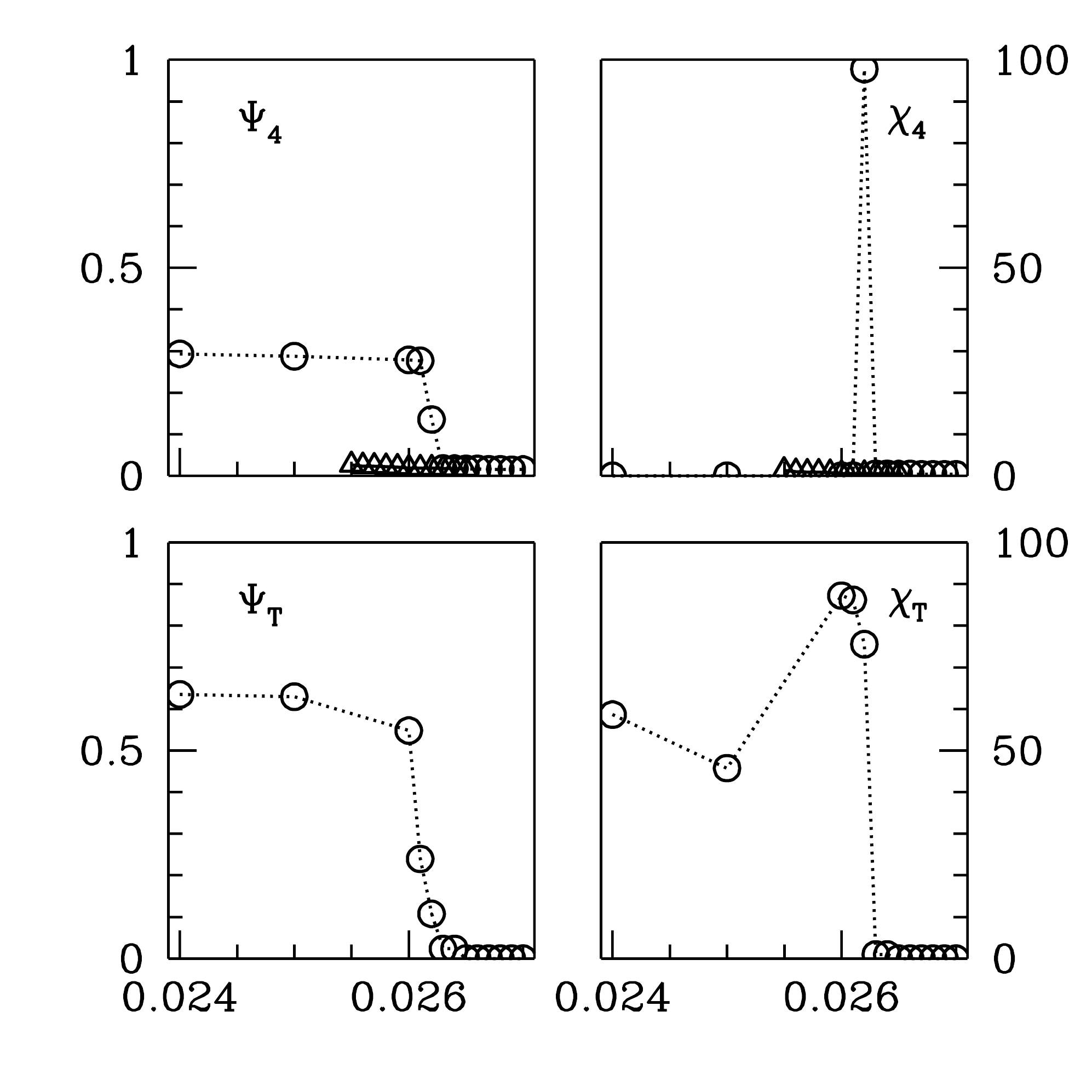}
\caption{YK potential with $a=3.3$ for $P=3$: order parameters and
susceptibilities in the $T$ range across the melting transition.
See the caption of Fig.\,9 for notation.}
\label{fig11}
\end{figure}

%
%
\begin{figure}
\includegraphics[width=9cm]{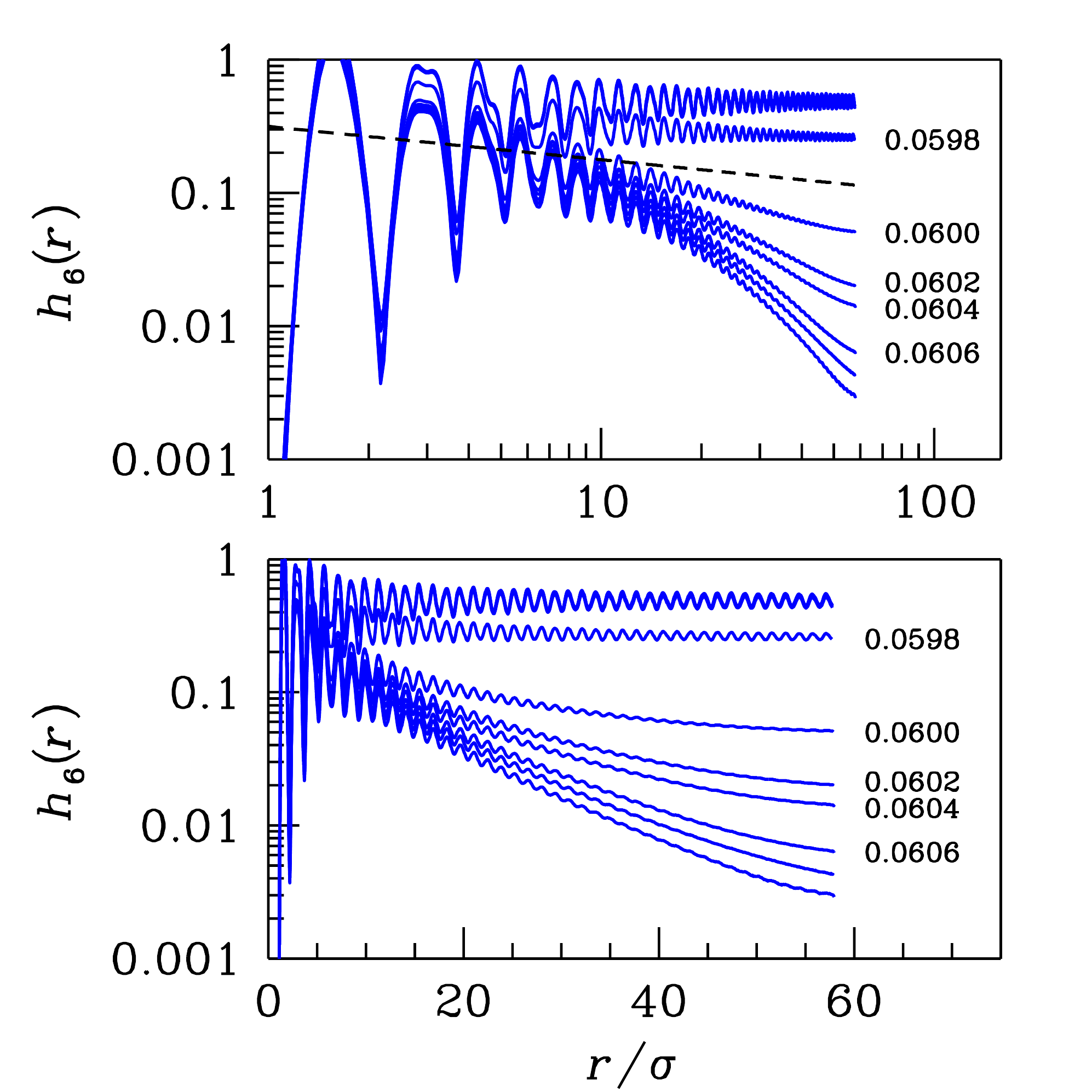}
\caption{(Color online). Orientational correlation function $h_6(r)$
at selected temperatures across the hexatic region for $P=0.5$ ($N=6048$).
Top: log-log plot; bottom: log-lin plot. Upon increasing $T$ from 0.0598
to 0.0606 there is a qualitative change in the large-distance behavior of
$h_6(r)$, from constant (triangular solid) to power-low decay (hexatic fluid),
up to
exponential decay (isotropic fluid). Note that, consistently with the KTHNY
theory, the decay exponent $\eta$ is less than $1/4$ (which is the slope of
the dashed straight line) in the hexatic phase.
The slight recovery of correlations which is observed near the largest
distance at which the OCF is computed (roughly corresponding to half
of the simulation-box length) is a finite-size effect due to the use
of periodic boundary conditions.}
\label{fig12}
\end{figure}

%
%
\begin{figure}
\includegraphics[width=9cm]{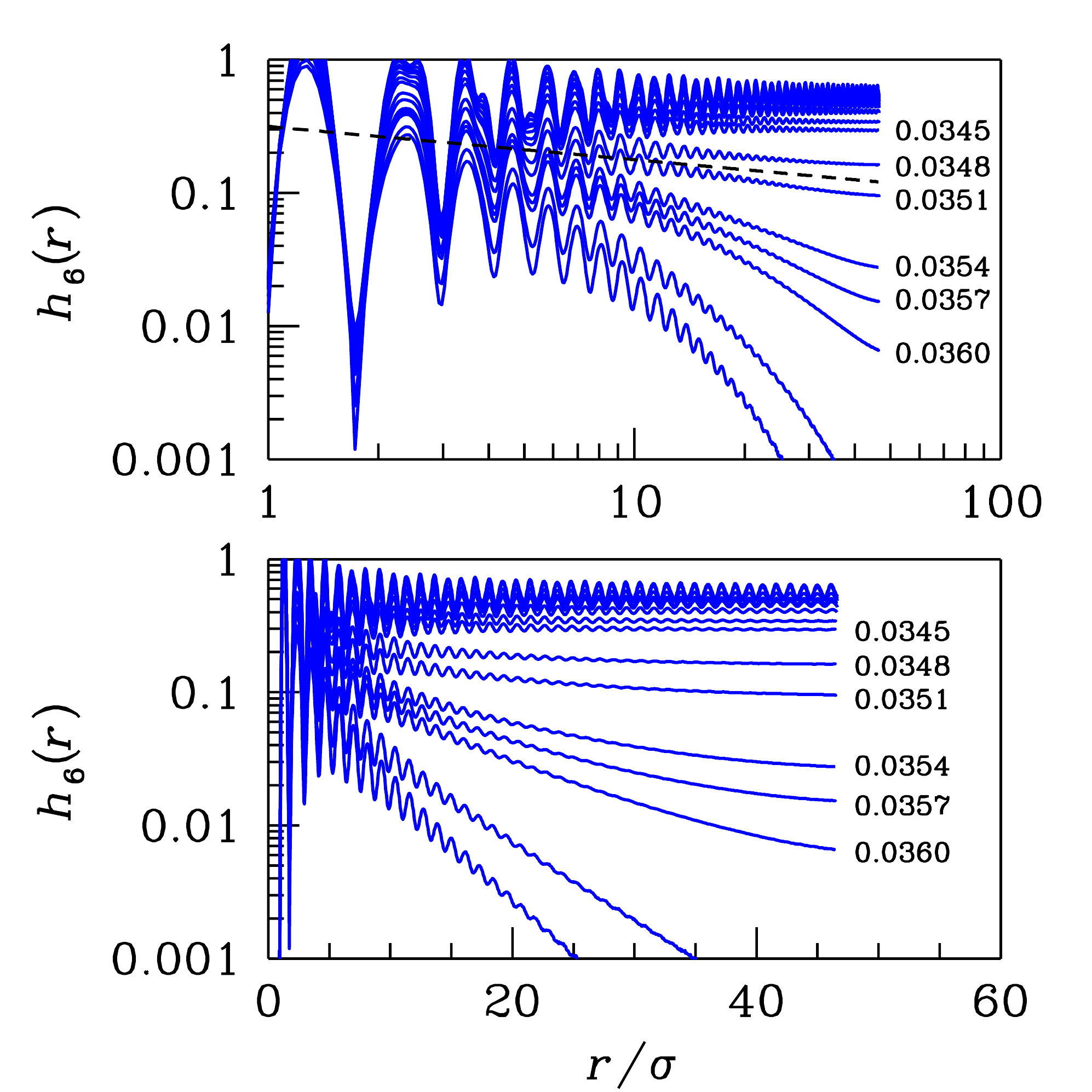}
\caption{(Color online). Orientational correlation function $h_6(r)$
at selected temperatures across the hexatic region for $P=2$ ($N=6048$).
Top: log-log plot; bottom: log-lin plot. Upon increasing $T$ from 0.0345
to 0.0360 there is a qualitative change in the large-distance behavior of
$h_6(r)$, from constant (triangular solid) to power-low decay (hexatic fluid),
up to
exponential decay (isotropic fluid). Moreover, the decay exponent $\eta$
is less than $1/4$ (which is the slope of the dashed straight line) in the
hexatic phase.}
\label{fig13}
\end{figure}

%
%
\begin{figure}
\includegraphics[width=9cm]{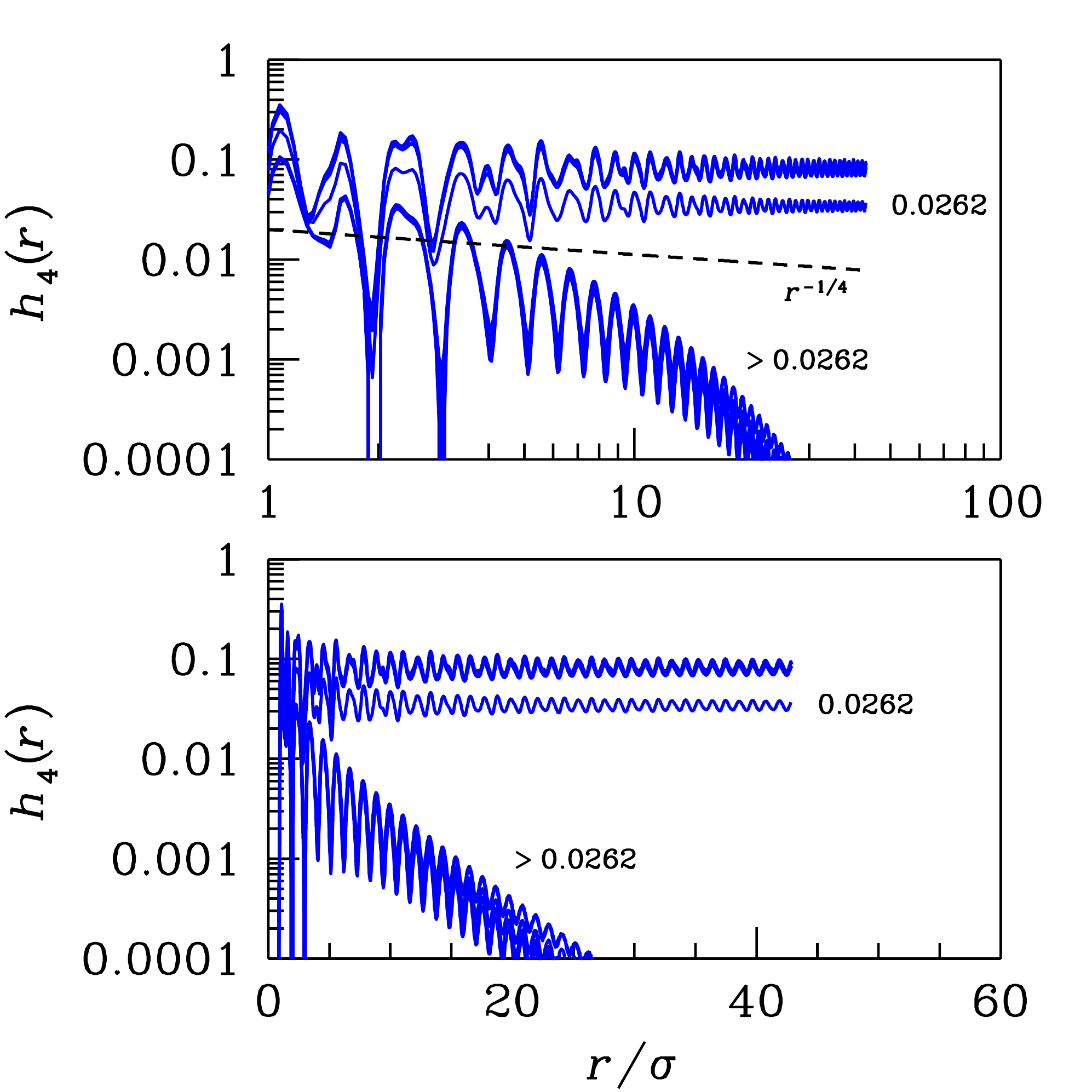}
\caption{(Color online). Orientational correlation function $h_4(r)$
at selected temperatures across the hexatic region for $P=3$ ($N=6084$).
Top: log-log plot; bottom: log-lin plot. At variance with the
triangular-lattice case, we assist to an abrupt change of decay mode as
$T$ goes from 0.0262 to 0.0263, from constant (square solid) directly to
exponential (isotropic fluid).}
\label{fig14}
\end{figure}

\begin{thebibliography}{99}
\bibitem{KTHNY} J. M. Kosterlitz and D. J. Thouless, {\em J. Phys. C} {\bf 6},
1181 (1973); B. I. Halperin and D. R. Nelson, {\em Phys. Rev. Lett.} {\bf 41},
121 (1978); A. P. Young, {\em Phys. Rev. B} {\bf 19}, 1855 (1979).

\bibitem{Chui} S. T. Chui, {\em Phys. Rev. B} {\bf 28}, 178 (1983).

\bibitem{Bernard} E. P. Bernard and W. Krauth, {\em Phys. Rev. Lett.}
{\bf 107}, 155704 (2011).

\bibitem{Chen} K. Chen, T. Kaplan, and M. Mostoller, {\em Phys. Rev. Lett.}
{\bf 74}, 4019 (1995).

\bibitem{Mejia} S. J. Mej\'ia-Rosales, A. Gil-Villegas, B. I. Ivlev, and
J. Ruiz-Garcia, {\em J. Phys. Chem. B} {\bf 110}, 22230 (2006).

\bibitem{Murray} C. A. Murray and D. H. Van Winkle, {\em Phys. Rev. Lett.}
{\bf 58}, 1200 (1987).

\bibitem{Marcus} A. H. Marcus and S. A. Rice, {\em Phys. Rev. Lett.}
{\bf 77}, 2577 (1996).

\bibitem{Zahn} K. Zahn, R. Lenke, and G. Maret, {\em Phys. Rev. Lett.}
{\bf 82}, 2721 (1999).

\bibitem{Keim} P. Keim, G. Maret, and H. H. von Gr\"{u}nberg,
{\em Phys Rev. E} {\bf 75}, 031402 (2007).

\bibitem{Lin1} B.-J. Lin and L.-J. Chen, {\em J. Chem. Phys} {\bf 126},
034706 (2007).

\bibitem{Han} Y. Han, N. Y. Ha, A. M. Alsayed, and A. G. Yodh,
{\em Phys. Rev. E} {\bf 77}, 041406 (2008).

\bibitem{Peng} Y. Peng, Z. Wang, A. M. Alsayed, A. G. Yodh, and Y. Han,
{\em Phys. Rev. Lett.} {\bf 104}, 205703 (2010).

\bibitem{Muto} S. Muto and H. Aoki, {\em Phys. Rev. B} {\bf 59}, 14911
(1999).

\bibitem{Lin2} S. Z. Lin, B. Zheng, and S. Trimper, {\em Phys Rev. E}
{\bf 73}, 066106 (2006).

\bibitem{Lee} S. I. Lee and S. J. Lee, {\em Phys. Rev. E} {\bf 78}, 041504
(2008).

\bibitem{Qi} W.-K. Qi, Z. Wang, Y. Han, and Y. Chen, {\em J. Chem. Phys.}
{\bf 133}, 234508 (2010).

\bibitem{Clark} B. K. Clark, M. Casula, and D. M. Ceperley,
{\em Phys. Rev. Lett.} {\bf 103}, 055701 (2009).

\bibitem{Prestipino1} S. Prestipino, F. Saija, and P. V. Giaquinta,
{\it Phys. Rev. Lett.} {\bf 106}, 235701 (2011).

\bibitem{Likos1} C. N. Likos, {\em Phys. Rep.} {\bf 348}, 267 (2001).

\bibitem{Stillinger} F. H. Stillinger, {\em J. Chem. Phys.} {\bf 65},
3968 (1976).

\bibitem{Lang} A. Lang, C. N. Likos, M. Watzlawek, and H. L\"owen,
{\em J. Phys.: Condens. Matter} {\bf 12}, 5087 (2000).

\bibitem{Prestipino2} S. Prestipino, F. Saija, and P. V. Giaquinta,
{\it Phys. Rev. E} {\bf 71}, 050102(R) (2005).

\bibitem{Mausbach} See e.g. P. Mausbach and H.-O. May,
{\em Fluid Phase Equilib.} {\bf 249}, 17 (2006).

\bibitem{Speranza} M. C. Speranza, S. Prestipino, and P. V. Giaquinta,
{\it Mol. Phys.} {\bf 109}, 3001 (2011).

\bibitem{Buldyrev} S. V. Buldyrev, G. Malescio, C. A. Angell,
N. Giovambattista, S. Prestipino, F. Saija, H. E. Stanley, L. Xu,
{\it J. Phys.: Condens. Matter} {\bf 21}, 504106 (2009).

\bibitem{Hemmer} P. C. Hemmer and G. Stell, {\it Phys. Rev. Lett.}
{\bf 24}, 1284 (1970)

\bibitem{Young} H. J. Young and B. J. Alder, {\it Phys. Rev. Lett.} {\bf 38},
1213 (1979); {\it J. Chem. Phys.} {\bf 70}, 473 (1979).

\bibitem{Debenedetti} P. G. Debenedetti, V. S. Raghavan, and S. S. Borick,
{\it J. Phys. Chem.} {\bf 95}, 4540 (1991).

\bibitem{Sadr} M. R. Sadr-Lahijany, A. Scala, S. Buldyrev, and H. E. Stanley,
{\it Phys. Rev. Lett.} {\bf 81}, 4895 (1998).

\bibitem{Jagla} E. A. Jagla, {\it J. Chem. Phys.} {\bf 111}, 8980 (1999).

\bibitem{Likos2} C. N. Likos, A. Lang, M. Watzlawek, and H. L\"owen,
{\it Phys. Rev. E} {\bf 63}, 031206 (2001).

\bibitem{Kumar} P. Kumar, S. V. Buldyrev, F. Sciortino, E. Zaccarelli,
H. E. Stanley, {\it Phys. Rev. E} {\bf 72}, 021501 (2005).

\bibitem{Deoliveira} A. B. de Oliveira, P. A. Netz, T. Colla, and
M. C. Barbosa, {\it J. Chem. Phys.} {\bf 124}, 084505 (2006).

\bibitem{Gibson} H. M. Gibson and N. B. Wilding, {\it Phys. Rev. E}
{\bf 73}, 061507 (2006).

\bibitem{Lomba} E. Lomba, N. G. Almarza, C. Martin, C. McBridge, 
{\it J. Chem. Phys.} {\bf 126}, 244510 (2006).

\bibitem{Fomin} D. Yu. Fomin, N. V. Gribova, V. N. Ryzhov, S. M. Stishov,
D. Frenkel, {\it J. Chem. Phys.} {\bf 129}, 064512 (2008).

\bibitem{Xu} L. Xu, S. Buldyrev, C. A. Angell, and H. E. Stanley,
{\it J. Chem. Phys.} {\bf 130}, 054505 (2009). 

\bibitem{Pizio} O. Pizio, H. Dominguez, Y. Duda, and S. Sokolowski,
{\it J. Chem. Phys.} {\bf 130}, 174504 (2009).

\bibitem{Pamies} J. C. P\`amies, A. Cacciuto, and D. Frenkel,
{\em J. Chem. Phys.} {\bf 131}, 044514 (2009).

\bibitem{Pauschenwein} G. J. Pauschenwein and G. Kahl, {\it J. Chem. Phys.}
{\bf 129}, 174107 (2008).

\bibitem{Malescio1} G. Malescio, F. Saija, and S. Prestipino,
{\it J. Chem. Phys.} {\bf 129}, 241101 (2008).

\bibitem{Scala} A. Scala, M. Reza Sadr-Lahijany, N. Giovambattista, 
S. V. Buldyrev, and H. E. Stanley, {\it Phys. Rev. E} {\bf 63}, 041202 (2001).

\bibitem{Wilding} N. B. Wilding and J. E Magee,
{\it Phys. Rev. E} {\bf 66}, 031509 (2002).

\bibitem{Almudallal} A. M. Almudallal, S. V. Buldyrev, and I. Saika-Voivod,
{\it J. Chem. Phys.} {\bf 137}, 034507 (2012).

\bibitem{Malescio2} G. Malescio and G. Pellicane,
{\it Nat. Mat.} {\bf 2}, 97 (2003).

\bibitem{Saija} F. Saija, S. Prestipino, and G. Malescio,
{\it Phys. Rev. E} {\bf 80}, 031502 (2009). 

\bibitem{Prestipino3} S. Prestipino, F. Saija, and G. Malescio,
{\it J. Chem. Phys.} {\bf 133}, 144504 (2010).

\bibitem{Malescio3} G. Malescio, S. Prestipino, and F. Saija,
{\it Mol. Phys.} {\bf 109}, 2837 (2011).

\bibitem{Malescio4} G. Malescio and F. Saija,
{\it J. Phys. Chem. B} {\bf 115}, 14091 (2011).

\bibitem{Dullens} R. P. A. Dullens and W. K. Kegel, {\em Phys. Rev. Lett.}
{\bf 92}, 195702 (2004).

\bibitem{Alsayed} A. M. Alsayed, Y. Han, and A. G. Yodh, {\it Melting and
geometric frustration in temperature-sensitive colloids}, A. Fernandez-Nieves
{\it et al.} eds. (Wiley-VCH, 2011). 

\bibitem{Yoshida} T. Yoshida and S. Kamakura, {\it Prog. Theor. Phys.}
{\bf 52}, 822 (1974).

\bibitem{Prestipino4} S. Prestipino, {\em J. Phys.: Condens. Matter} {\bf 24},
035102 (2012).

\bibitem{Prestipino5} S. Prestipino, F. Saija, and G. Malescio,
{\it Soft Matter} {\bf 5}, 2795 (2009).

\bibitem{Frenkel} D. Frenkel and B. Smit, {\em Understanding molecular
simulation}, 2nd ed. (Academic Press, 2002).

\bibitem{Prestipino6} See e.g. S. Prestipino and P. V. Giaquinta,
{\em J. Stat. Phys.} {\bf 96}, 135 (1999).

\bibitem{Jabes} B. S. Jabes, M. Agarwal, and C. Chakravarty,
{\em J. Chem. Phys.} {\bf 132}, 234507 (2010).
\end{thebibliography}
\end{document}